\documentclass{article}

\usepackage[super]{cite}
\usepackage{xcolor}
\usepackage{longtable}
\usepackage[verbose,hypertexnames=false]{hyperref}
\hypersetup{colorlinks=false,allbordercolors=blue,pdfborderstyle={/S/U/W 1}}

\begin{document}



\title{Exact quasinormal modes in Grumiller spacetime }

\author{Li-Qin Mi
\\\textsl{School of Physics,}\\
\textsl{Zhejiang University of Technology, Hangzhou 310023, China}\\
\textsl{lqmi@zjut.edu.cn}\\\\
Zhong-Heng Li\footnote{Corresponding authors.}
\\\textsl{Black Hole and Gravitational Wave Group,}\\
\textsl{Zhejiang University of Technology,}\\
\textsl{Hangzhou 310023, China}\\
\textsl{zhli@zjut.edu.cn}}

\maketitle


\begin{abstract}
The Grumiller metric is an effective model for gravity at large distances and plays a significant role in constructing galactic models and explaining dark matter. Here, in Grumiller spacetime, we analytically compute the quasinormal-mode frequencies and wave functions for massless particles with spin $\leq 2$ by introducing a new transformation relation. Our findings indicate that the quasinormal-mode frequencies are identical for different fermions with the same quantum number $n$. Notably, no bosonic quasinormal modes associated with Heun polynomials were found. Furthermore, for a given quasinormal-mode frequency, the corresponding particles exhibit a $2(n + 1)$-fold degeneracy. These results provide a theoretical basis for the mutual simulation of fermionic waves.
\end{abstract}



\section{Introduction}
The first direct detection of gravitational waves was achieved in 2015.\cite{Abbott} During the subsequent decade, the LIGO Scientific Collaboration has identified more than 100 gravitational-wave events originating from binary black hole mergers. In the post-merger phase, the distorted remnant black hole evolves toward equilibrium through gravitational-wave emission. This ringdown signal is spectrally dominated by exponentially damped sinusoids, making quasinormal modes the characteristic fingerprints of a perturbed black hole's final state.

Since Vishveshwara\cite{Vishveshwara} first identified quasinormal modes in perturbations of Schwarzschild spacetime, extensive research has been conducted over the past fifty-five years on quasinormal modes and their associated frequencies for various black hole types.\cite{Nollert,Kokkotas,Berti,Konoplya} Notably, in recent years, the high-precision observational capabilities of future space-based gravitational wave detectors -- such as LISA,\cite{Danzmann,LISA} Taiji,\cite{Hu} and TianQin\cite{TianQin} -- have motivated growing efforts to determine quasinormal modes and frequencies with higher accuracy.\cite{Bourg}

Despite widespread interest, studies remain lacking in two key areas: quasinormal modes for gravity at large distance, and analogies between quasinormal-mode frequencies of massless spin-particle waves. Due to dark energy\cite{Carroll} and dark matter,\cite{Bertone} gravity at large distance constitutes one of the most critical research frontiers in modern gravitational physics. In 2010, Grumiller\cite{Grumiller} proposed a solution of the Einstein field equations with an anisotropic fluid. This solution effectively models gravity at large distances. The aim of this paper is to build on this result in order to analytically study its quasinormal modes and explore the analogy between massless spin-particle waves.

\section{Wave equation and its solutions}

The spacetime line element proposed by Grumiller\cite{Grumiller} takes the form
\begin{equation}
ds=K^{2}dt^{2}-\frac{dr^{2}}{K^{2}}-r^{2}(d\theta^{2}+\sin^{2}\theta d\phi^{2}),
\label{eq:1}
\end{equation}
where
\begin{equation}
K^{2}=1-\frac{2M}{r}-\Lambda r^{2}+2\textsl{a} r.
\label{eq:2}
\end{equation}
Here $\Lambda$ denotes the cosmological constant, $M$ represents the black hole mass, and a is the Rindler acceleration. While the $dt$ and $dr$ components of this line element originate from the action
\begin{equation}
\textbf{S}=-\int d^{2}x \sqrt{-g}[\Phi^{2}R+2 (\partial \Phi)^{2}-6\Lambda \Phi^{2}+8 a \Phi+2].
\label{eq:3}
\end{equation}
The complete line element can be interpreted as an exact solution of the Einstein field equations with a cosmological constant. The associated effective energy-momentum tensor describes an anisotropic fluid:
\begin{equation}
T^{\mu}_{\nu}=diag(-\rho, p_{r}, p_{\bot}, p_{\bot}),
\label{eq:4}
\end{equation}
with density $\rho$, and $p_{r}$ and tangential $p_{\bot}$ pressure,
\begin{equation}
\rho=\frac{a}{2\pi r}, \quad  p_{r}=-\rho,  \quad  p_{\bot}=\frac{1}{2} p_{r}.
\label{eq:5}
\end{equation}
Notably, this line element exhibits the specific signature required to explain both galactic rotation curves and the Pioneer anomaly. Furthermore, this framework predicts a distinctive yet physically plausible equation of state (5) for dark matter. This prediction could be empirically tested through combined analyses of gravitational lensing observations and rotation curve measurements.\cite{Grumiller}

The event horizon equation for the spacetime defined by metric (1) is
\begin{equation}
1-\frac{2M}{r}-\Lambda r^{2}+2\textsl{a} r=0.
\label{eq:6}
\end{equation}
This equation admits three distinct real roots, with two positive roots and one negative root. Denoting the positive roots as $r_{c}$ and $r_{b}$ ($r_{c} > r_{b}$) and the negative root as $r_{n}$, we identify $r_{c}$ as the cosmological horizon and $r_{b}$ as the black hole horizon. These roots can be expressed as follows:
\begin{eqnarray}\large
r_{c}&=&\frac{2}{3 \Lambda}(3\Lambda+4\textsl{a}^{2})^{\frac{1}{2}}\cos(\frac{\vartheta}{3}),\nonumber\\
r_{n}&=&\frac{2}{3 \Lambda}(3\Lambda+4\textsl{a}^{2})^{\frac{1}{2}}\cos(\frac{\vartheta}{3}+\frac{2\pi}{3}),\nonumber\\
r_{b}&=&\frac{2}{3 \Lambda}(3\Lambda+4\textsl{a}^{2})^{\frac{1}{2}}\cos(\frac{\vartheta}{3}+\frac{4\pi}{3}),
\label{eq:7}
\end{eqnarray}
where
\begin{equation}
\cos\vartheta=\frac{9\textsl{a}\Lambda-27M\Lambda^{2}+8\textsl{a}^{3}}{(3\Lambda+4\textsl{a}^{2})^{\frac{3}{2}}}.
\label{eq:8}
\end{equation}

It is known that all spherically symmetric spacetimes are algebraically special of type D,\cite{Li_a} and thus the metric (1) therefore belongs to this class. Crucially, while wave equations for massless fields -- including the Weyl neutrino ($s=1/2, p=\pm 1/2$), electromagnetic ($s=1, p=\pm 1$), massless Rarita-Schwinger ($s=3/2, p\pm 3/2$), and gravitational ($s=2, p=\pm 2$) fields -- generally resist exact decoupling, they admit full decoupling under perturbations in all type-D metrics.\cite{Teukolsky,Torres}

This implies that each spin state $p$ corresponds to a distinct field equation. Using $\Phi_{p}$ to denote the wave function for a given spin state, our recent work\cite{Li_b} demonstrates that the decoupled equations for spins $0, 1/2, 1, 3/2$, and $2$ in spacetime (1) can be unified,  i.e., through the transformation:
\begin{equation}
\Phi_{p}=r^{(p-s)}\Psi_{p}.
\label{eq:9}
\end{equation}
all equations reduce to a single elegant form (source-free case):\cite{Li_b}
\begin{equation}
[(\nabla^{\mu}+pL^{\mu})(\nabla_{\mu}+pL_{\mu})-4p^{2}\psi_{2}+\frac{1}{6}R]\Psi_{p}=0,
\label{eq:10}
\end{equation}
where
\begin{eqnarray}\large
L^{t}&=&\frac{3M-r(1+\textsl{a}r)}{r[2M-r(1+2\textsl{a}r-\Lambda r^{2})]},\nonumber\\
L^{r}&=&\frac{M}{r^{2}}-\frac{1}{r}+2\Lambda r-3\textsl{a},\nonumber\\
L^{\theta}&=&0,\nonumber\\
L^{\varphi}&=&-\frac{1}{r^{2}}\frac{i \cos\theta}{\sin^{2}\theta}.
\label{eq:11}
\end{eqnarray}
\begin{equation}
\psi_{2}=-\frac{M}{r^{3}}.
\label{eq:12}
\end{equation}
\begin{equation}
R=12(-\frac{\textsl{a}}{r}+\Lambda).
\label{eq:13}
\end{equation}
Here, $\nabla_{\mu}$, $\psi_{2}$, and $R$ are the covariant derivative, Weyl scalar, and scalar curvature, respectively.

Equation (10) can be solved using the method of separation of variables; the solution takes the form
\begin{equation}
\Psi_{p}=r^{-(2p+1)} e^{-i\omega t} e^{i \omega r_{*}} S(\theta, \varphi) y(r),
\label{eq:14}
\end{equation}
where $r_{*}$ is called the tortoise coordinate. It is determined by the equation:
\begin{equation}
\partial^{\mu}v \partial_{\mu}v=0, \quad \partial^{\mu}u \partial_{\mu}u=0.
\label{eq:15}
\end{equation}
Here $v$ and $u$ are the Eddington-Finkelstein null coordinates, which take the form
\begin{equation}
v=t+r_{*}, u=t-r_{*}.
\label{eq:16}
\end{equation}
By substituting Eq. (16) into Eq. (15) and using metric (1), we derive the exact form of the tortoise coordinate:
\begin{equation}
r_{*}=\frac{1}{2\kappa_{b}}\ln|\frac{r}{r_{b}}-1|+\frac{1}{2\kappa_{c}}\ln|\frac{r}{r_{c}}-1|+\frac{1}{2\kappa_{n}}\ln|\frac{r}{r_{n}}-1|.
\label{eq:17}
\end{equation}
with
\begin{equation}
\kappa_{b}=-\frac{\Lambda}{2 r_{b}}(r_{b}-r_{c})(r_{b}-r_{n}),
\label{eq:18}
\end{equation}
\begin{equation}
\kappa_{c}=-\frac{\Lambda}{2 r_{c}}(r_{c}-r_{b})(r_{c}-r_{n}),
\label{eq:19}
\end{equation}
\begin{equation}
\kappa_{n}=-\frac{\Lambda}{2 r_{n}}(r_{n}-r_{b})(r_{n}-r_{c}).
\label{eq:20}
\end{equation}
where $\kappa_{b}$ denotes the surface gravity of the black hole horizon, while $\kappa_{c}$ represents the surface gravity of the cosmological horizon.

By substituting equation (14) into equation (10) and using the variable transformation
\begin{equation}
z=\frac{r_{c}}{r_{c}-r_{b}}\frac{r-r_{b}}{r},
\label{eq:21}
\end{equation}
we decompose equation (10) into the transverse and radial equations:
\begin{eqnarray}
\big[\frac{1}{\sin\theta}\frac{\partial}{\partial\theta}\big(\sin\theta\frac{\partial}{\partial\theta}\big)&+&\frac{1}{\sin^{2}\theta}\frac{\partial^{2}}{\partial\varphi^{2}}
+\frac{2ip\cos\theta}{\sin^{2}\theta}\frac{\partial}{\partial\varphi}\nonumber\\
&-&p^{2}\cot^{2}\theta+p+\lambda \big]S(\theta,\varphi)=0,
\label{eq:22}
\end{eqnarray}
and
\begin{eqnarray}
\frac{d^{2}y}{dz^{2}}+(\frac{\gamma}{z}+\frac{\delta}{z-1}+\frac{\epsilon}{z-a})\frac{dy}{dz}+\frac{\alpha\beta z-q}{z(z-1)(z-a)}y=0,
\label{eq:23}
\end{eqnarray}
with
\begin{equation}
\gamma+\delta+\epsilon=\alpha+\beta+1.
\label{eq:24}
\end{equation}
Here,
\begin{eqnarray}
&&\gamma=\frac{i \omega}{\kappa_{b}}+p+1, \quad \delta=\frac{i \omega}{\kappa_{c}}+p+1, \quad \epsilon=\frac{i \omega}{\kappa_{n}}+p+1, \nonumber\\
&&\alpha=p+1, \quad \beta=2p+1,  \quad a=\frac{(r_{n}-r_{b}) r_{c}}{(r_{c}-r_{b}) r_{n}}, \nonumber\\
&&q=\frac{(2p+1)(p+1)r_{c} r_{n} \Lambda-\lambda}{(r_{c}-r_{b}) r_{n} \Lambda}.
\label{eq:25}
\end{eqnarray}
where $\lambda$ is the constant arising from the separation of variables.

The radial equation (23) is a standard Heun equation,\cite{Ronveaux} with regular singularities at $0, 1, a, \infty$ and corresponding exponents $\{0, 1-\gamma\}$, $\{0, 1-\delta\}$, $\{0, 1-\epsilon\}$, and $\{\alpha, \beta\}$, respectively. Here, $a$ is known as the singularity parameter; $\alpha$, $\beta$, $\gamma$, and $\epsilon$ are referred to as exponent parameters; and $q$ is called the accessory parameter. Due to relation (24), the total number of free parameters is six.

According to the theory of Heun equation, Eq. (23) has two solutions at $z=0$ corresponding to the two exponents at that point. The solution corresponding to the exponent $0$ is\cite{Ronveaux}
\begin{equation}
H\ell(a,q;\alpha,\beta,\gamma,\delta;z),
\label{eq:26}
\end{equation}
and the solution corresponding to the exponent $1-\gamma$ is
\begin{equation}
z^{1-\gamma} H\ell(a,(a\delta+\epsilon)(1-\gamma)+q;\alpha+1-\gamma,\beta+1-\gamma,2-\gamma,\delta;z).
\label{eq:27}
\end{equation}

Note that only when $\gamma \notin \{0, -1, -2, \ldots\}$ does $H\ell(a,q;\alpha,\beta,\gamma,\delta;z)$ exist, be analytic in the disk $|z|<1$, and admit the Maclaurin expansion\cite{Ronveaux}
\begin{equation}
H\ell(a,q;\alpha,\beta,\gamma,\delta;z)=\sum^{\infty}_{j=0}c_{j}z^{j}, \quad |z|<1,
\label{eq:28}
\end{equation}
where $c_{0}=1$,
\begin{eqnarray}
&&a\gamma c_{1}=q c_{0}, \nonumber\\
&&a(j+1)(j+\gamma)c_{j+1}-j[(j-1+\gamma)(1+a)+a\delta+\epsilon]c_{j} \nonumber\\
&&+(j-1+\alpha)(j-1+\beta)c_{j-1}=q c_{j}, \quad j\geq 1.
\label{eq:29}
\end{eqnarray}

\section{Frequencies and radial wave functions of quasinormal modes }

It is widely known that the quasinormal-mode boundary conditions require the wave to be purely ingoing at the event horizon and purely outgoing at spatial infinity; therefore, the radial wave function $R_{p}$ must satisfy\cite{Nollert,Kokkotas,Berti,Konoplya}
\begin{equation}
R_{p}\sim\
\left\{
\begin{array}{ c c }
e^{-i\omega r_{\ast}}, & r_{\ast}\rightarrow -\infty; \\
e^{i\omega r_{\ast}}, & r_{\ast}\rightarrow \infty. \\
\end{array}
\right.
\label{eq:30}
\end{equation}

 For the quasinormal modes in the spacetime described by metric (1), it is natural to assume that they satisfy the boundary conditions (30). To obtain the radial wave function that satisfies these conditions, we employ Equation (23), whose general solution, as discussed previously, is given by
\begin{eqnarray}
y=&&D_{1} H\ell(a,q;\alpha,\beta,\gamma,\delta;z) \nonumber\\
+&&D_{2} z^{1-\gamma} H\ell(a,(a\delta+\epsilon)(1-\gamma)+q;\alpha+1-\gamma,\beta+1-\gamma,2-\gamma,\delta;z).
\label{eq:31}
\end{eqnarray}
Here, $D_{1}$ and $D_{2}$ are arbitrary constants. Based on Eqs. (9), (14), and (31), the wave function $\Phi_{p}$ for all massless spin particles takes the form
\begin{eqnarray}
\Phi_{p}=&&r^{(p-s)}\Psi_{p}=e^{-i\omega t} e^{i \omega r_{*}} r^{-(s+p+1)} S(\theta,\varphi) y \nonumber\\
=&&e^{-i\omega t} e^{i \omega r_{*}} r^{-(s+p+1)} S(\theta,\varphi)[D_{1} H\ell(a,q;\alpha,\beta,\gamma,\delta;z) \nonumber\\
+&&D_{2} z^{1-\gamma} H\ell(a,(a\delta+\epsilon)(1-\gamma)+q;\alpha+1-\gamma,\beta+1-\gamma,2-\gamma,\delta;z)].
\label{eq:32}
\end{eqnarray}
Therefore, the specific expression for the radial wave function $R_{p}$ of massless particles with arbitrary spin is
\begin{eqnarray}
R_{p}=&&e^{i \omega r_{*}} r^{-(s+p+1)} y \nonumber\\
=&&e^{i \omega r_{*}} r^{-(s+p+1)}[D_{1} H\ell(a,q;\alpha,\beta,\gamma,\delta;z) \nonumber\\
+&&D_{2} z^{1-\gamma} H\ell(a,(a\delta+\epsilon)(1-\gamma)+q;\alpha+1-\gamma,\beta+1-\gamma,2-\gamma,\delta;z)].
\label{eq:33}
\end{eqnarray}

As $r_{*}\rightarrow -\infty$ ($r\rightarrow r_{b}$) near the event horizon, the radial wave function has the asymptotic behavior:

\begin{equation}
R_{p}=D_{1} e^{i \omega r_{*}}+D_{2} e^{-(2\kappa_{b} p+i \omega)r_{*}}.
\label{eq:34}
\end{equation}
The boundary condition at $r=r_{b}$ requires that $D_{1}=0$, and hence we have
\begin{eqnarray}
R_{p}=&&D_{2} e^{i \omega r_{*}} r^{-(s+p+1)} z^{1-\gamma} \nonumber\\
\cdot &&H\ell[(a,(a\delta+\epsilon)(1-\gamma)+q;\alpha+1-\gamma,\beta+1-\gamma,2-\gamma,\delta;z)].
\label{eq:35}
\end{eqnarray}

To obtain the quasinormal modes, it is necessary to analyze the behavior of the radial wave function given in equation (35) as $r_{*}\rightarrow\infty$. Note that in this limit, $r\rightarrow r_{c}$ and $z\rightarrow 1$, and the radial wave function behaves as:
\begin{equation}
R_{p}\sim e^{i \omega r_{*}} H\ell[(a,(a\delta+\epsilon)(1-\gamma)+q;\alpha+1-\gamma,\beta+1-\gamma,2-\gamma,\delta;z)].
\label{eq:36}
\end{equation}
Similar to Equation (28), if $2-\gamma \neq 0, -1, -2, ...$, the Heun function exists and is analytic in the disk $\mid z \mid<1$, admitting a Maclaurin expansion. However, this still does not guarantee that the series converges at $z=1$, i.e., the radial wave function (35) does not satisfy the boundary condition at infinity.

There is only one way to resolve this issue: the series for quasinormal modes must be forcibly truncated, becoming Heun polynomials. In that case, as $z=1$, $H\ell[(a,(a\delta+\epsilon)(1-\gamma)+q;\alpha+1-\gamma,\beta+1-\gamma,2-\gamma,\delta;z)]$ is a finite constant. Consequently, the radial wave function satisfies the boundary condition required for quasinormal modes as $r_{*}\rightarrow\infty$.

The condition for Heun series in Eq. (36) to become a polynomial of degree $n$ is
\begin{equation}
\alpha+1-\gamma==-n, \quad (n=0, 1, 2, ...),
\label{eq:37}
\end{equation}
This gives remarkably simple formulae for the frequencies of the quasinormal modes:
\begin{equation}
\omega=-i \kappa_{b} (n+1),
\label{eq:38}
\end{equation}
With $\alpha=p + 1$, Equation (37) gives  $2-\gamma=-(n+p)$, implying that $2-\gamma$ is a negative integer for bosons and thereby violating the analyticity condition for the Heun function in Equation (36). We therefore conclude that no bosonic quasinormal modes described by Heun polynomials exist in the spacetime of metric (1).

Equation (38) reveals that different fermions can share identical quasinormal-mode frequencies at the same quantum number $n$. As these frequencies are determined solely by the black hole's surface gravity and are independent of particle properties, this finding suggests that one type of fermion can emulate the quasinormal modes of another.

As mentioned above, the boundary conditions for quasinormal modes require that $H\ell[(a,(a\delta+\epsilon)(1-\gamma)+q;\alpha+1-\gamma,\beta+1-\gamma,2-\gamma,\delta;z)]$ be the Heun polynomial. For the polynomial of degree $n$, the coefficients $c_{j}$, which are obtained using the recurrence relation (29), can be written in the form of the following matrix equation:
\begin{equation}
\begin{array}{ll}
\left(\begin{array}{lll}
\,\,0\,\,\,\,\,\,a(2-\gamma)\,\,\,\,\,\,\,0\,\,\,\,\,\,\,\,\,\cdots\,\,\,\,\,\,\,\,0 \\
A_{1}\,\,\,\,\,\,-B_{1}\,\,\,\,\,\,\,\,\,\,C_{1}\,\,\,\,\,\,\,\cdots\,\,\,\,\,\,\,\,0 \\
\,\,0\,\,\,\,\,\,\,\,\,\,\,\,\,A_{2}\,\,\,\,\,\,-B_{2}\,\,\,\,\,\,\,\,\,\,\,\,\,\,\,\,\,\,\,\,\,\,\,\,\vdots \\
\,\,\,\vdots\,\,\,\,\,\,\,\,\,\,\,\,\,\,\,\,\vdots\,\,\,\,\,\,\,\,\,\,\,\,\,\,\,\,\,\,\vdots\,\,\,\,\,\,\,\,\,\,\ddots\,\,\,\,\,C_{n-1} \\
\,\,0\,\,\,\,\,\,\,\,\,\,\,\,\,\,\,0\,\,\,\,\,\,\,\,\,\,\,\,\,\,\cdots\,\,\,\,\,\,\,A_{n}\,\,\,\,-B_{n}
\end{array}
\right)
\left(\begin{array}{lll}
\,\,\,\,c_{0}\\
\,\,\,\,c_{1}\\
\,\,\,\,\,\vdots \\
c_{n-1}\\
\,\,\,\,c_{n}
\end{array}
\right)
=Q\left(\begin{array}{lll}
\,\,\,\,c_{0}\\
\,\,\,\,c_{1}\\
\,\,\,\,\,\vdots \\
c_{n-1}\\
\,\,\,\,c_{n}
\end{array}
\right),
\end{array}
\label{eq:39}
\end{equation}
with $c_{0}=1$,
\begin{eqnarray}
&&A_{j}=(j+\alpha-\gamma)(j+\beta-\gamma)=(j-1-n)(j-1-n+p), \nonumber\\
&&B_{j}=j[(j+1-\gamma)(1+a)+a\delta+\epsilon]=j\{(j-1-n-p)(1+\frac{1-r_{b}/r_{n}}{1-r_{b}/r_{c}}) \nonumber\\
&&\quad\quad+(\frac{1-r_{b}/r_{n}}{1-r_{b}/r_{c}})[1+p+(1+n)\frac{\kappa_{b}}{\kappa_{c}}]+1+p+(1+n)\frac{\kappa_{b}}{\kappa_{n}}\},  \nonumber\\
&&C_{j}=a(j+1)(j+2-\gamma)=(\frac{1-r_{b}/r_{n}}{1-r_{b}/r_{c}})(j+1)(j-n-p), \nonumber\\
&&Q=(a\delta+\epsilon)(1-\gamma)+q, \quad\quad (p=\pm \frac{1}{2}, \pm \frac{3}{2}).
\label{eq:40}
\end{eqnarray}
Note that the diagonal elements, $B_{j}$, have the following properties: $B_{j}(p) = B_{j}(-p)$ and $B_{j}(p) = B_{j}(p+1)$. Algebraic theory states that a necessary condition for the existence of a non-trivial solution is that $Q$ must be an eigenvalue of the tridiagonal matrix in Eq. (39). That is, $Q$ must take the discrete values $Q=Q_{n,m}$ with $m=0, 1, 2,...,n$. We make the ansatz that the eigenvector corresponding to $Q_{n,m}$ is $(c^{0}_{n,m}\,\,c^{1}_{n,m}\,\,...\,\,c^{n}_{n,m})^{T}$. Consequently, the final expression for the quasinormal-mode radial wave function is given by
\begin{eqnarray}
&&R_{p}=D_{2} e^{i \omega r_{*}} r^{-(s+p+1)} z^{1-\gamma} H\ell(a,Q_{n,m};-n,\beta+1-\gamma,2-\gamma,\delta;z) \nonumber\\
&&\quad \, \, \,=D_{2} e^{i \omega r_{*}} r^{-(s+p+1)} z^{1-\gamma} \sum^{n}_{j=0}c^{j}_{n,m} z^{j}.
\label{eq:41}
\end{eqnarray}

With the parameters $a=(1-r_{b}/r_{n})/(1-r_{b}/r_{c})=1.2956444250, \kappa_{b}/\kappa_{c}=-4.3824415940, \kappa_{b}/\kappa_{n}=3.3824415940$, we present in Table 1 and Table 2 the $Q_{n,m}$ and $c^{j}_{n,m}$ values, respectively, for the Weyl neutrino and massless Rarita-Schwinger field at quantum numbers $n=1$ and $n=2$.

\setlength{\tabcolsep}{4pt}
\renewcommand*{\arraystretch}{0.97}
\begin{longtable}{c|c|c|c|c}
\caption{Some eigenvalues of the tridiagonal matrix.}
\label{tab1} \\ \hline
       \multicolumn{1}{c}{$\rm $} \vline & \multicolumn{2}{c}{$\rm neutrino$} \vline &\multicolumn{2}{c}{$\rm  Rarita-Schwinger$} \\ \hline
        $\rm p$  & $\rm 1/2$ & $\rm -1/2$ & $\rm 3/2$ & $\rm -3/2$  \\ \hline
        $Q_{1,0}$     & 0.222422293  & 0.222422293   & -0.329148683  & -0.329148683    \\
        $Q_{1,1}$     & 4.368866557  & 4.368866557   & 4.920437532  & 4.920437532  \\
        $Q_{2,0}$     & 1.193211727  & 1.193211727  & 0.539991528 & 0.539991528    \\
        $Q_{2,1}$     & 8.384154135  & 8.384154135   & 8.053263845 & 8.053263845    \\
        $Q_{2,2}$     & 13.379078388  & 13.379078388   & 14.363188876  & 14.363188876   \\
        \hline
\end{longtable}

\setlength{\tabcolsep}{4pt}
\renewcommand*{\arraystretch}{0.97}
\begin{longtable}{c|c|c|c|c}
\caption{Some eigenvectors of the tridiagonal matrix.}
\label{tab1} \\ \hline
        \multicolumn{1}{c}{$\rm $} \vline & \multicolumn{2}{c}{$\rm neutrino$} \vline &\multicolumn{2}{c}{$\rm  Rarita-Schwinger$} \\ \hline
        $\rm p$  & $\rm 1/2$ & $\rm -1/2$ & $\rm 3/2$ & $\rm -3/2$  \\ \hline
        $c^{(0)}_{1,0}$     & 1  & 1   & 1  & 1    \\
        $c^{(1)}_{1,0}$     & -0.114446160  & -0.34333848   & 0.101616980  & -0.508084898  \\
        $c^{(0)}_{1,1}$     & 1  & 1  & 1 & 1    \\
        $c^{(1)}_{1,1}$     & -2.247976102  & -6.743928307   & -1.519070337 & 7.595351684    \\
        $c^{(0)}_{2,0}$     & 1  & 1   & 1  & 1   \\
        $c^{(1)}_{2,0}$     & -0.368376293  & -0.613960489   & -0.119078422  & -0.833548955   \\
        $c^{(2)}_{2,0}$     & 0.014640585  & 0.073202925   & -0.004499001  & 0.157465020   \\
        $c^{(0)}_{2,1}$     & 1  & 1   & 1  & 1   \\
        $c^{(1)}_{2,1}$     & -2.588412059  & -4.314020095   & -1.775898142  & -12.43128700   \\
        $c^{(2)}_{2,1}$     & 0.240125248  & 1.200626239   & -0.155219496  & 5.432682375   \\
        $c^{(0)}_{2,2}$     & 1  & 1   & 1  & 1   \\
        $c^{(1)}_{2,2}$     & -4.130478434  & -6.884130723   & -3.167356854  & -22.17149798  \\
        $c^{(2)}_{2,2}$     & 5.231259235  & 26.15629618   & 2.687287342  & -94.05505699   \\
        \hline
\end{longtable}

\section{Discussion and conclusion}

We have employed the unified wave equation\cite{Li_b} for massless spin particles to investigate the quasinormal modes in the Grumiller spacetime. We have identified a new transformation (14) that has mapped the radial part of the wave equation into a Heun equation, thereby providing a rigorous analytic framework for studying quasinormal modes. The essential feature of this transformation has been the introduction of the exponential factor $e^{i \omega r_{*}}$, and we anticipate that it can be generalized to other black hole backgrounds as a powerful tool for analyzing particle wave equations in curved spacetime.

We have derived the quasinormal-mode frequencies for fermions, as given by Eq. (38). These frequencies are found to take only discrete imaginary values. Recalling that $\Phi_{p}\sim e^{-i\omega t}$, a purely imaginary frequency $\omega$ corresponds directly to a damping rate. According to Equation (38), the quasinormal-mode frequency for fermions is given by $(n+1)$ times the surface gravity of the black hole. In principle, observation of the damping factors of these modes allows for the determination of the black hole's parameters. It is particularly noteworthy that the quasinormal-mode frequencies are independent of the particles' intrinsic angular momentum and their so-called ``extrinsic'' angular momentum, being determined exclusively by the black hole's properties. This feature establishes a theoretical basis for the simulation of one fermion wave type by another. It was also found that no quasinormal modes linked to Heun polynomials exist for bosons.

We have analyzed the  radial wave functions of the quasinormal modes, which contained both the exponential factor $e^{i\omega r_{*}}$ and Heun polynomials, as shown in Eq. (41). Since $Q_{n,m}$ appears as the eigenvalue of a tridiagonal matrix in Eq. (39), a generic $(n+1)$-dimensional matrix admits $n+1$ distinct eigenvalues. Each eigenvalue corresponds to an eigenvector. Equation (39) reveals that different fermion spin states are described by distinct tridiagonal matrices. Although the eigenvalues for the states $p=\pm s$ are identical (see Table 1), their eigenvectors differ (see Table 2). Since the quasinormal-mode frequency given by (38) is independent of the spin state $p$, the system exhibits a $2(n+1)$-fold degeneracy for a given frequency $\omega_{n}$, i.e., $2(n+ 1)$ linearly independent solutions of the wave equation share the same quasinormal-mode frequency $\omega_{n}$.

\section*{Acknowledgments}
This work was supported by the National Natural Science Foundation of China under Grant No. 12175198.


\begin{thebibliography}{0}

\bibitem{Abbott} B.P. Abbott et al. (LIGO Scientific, Virgo), Observation of gravitational waves from a
binary black hole merger, \textsl{Phys. Rev. Lett.} {\bf116}, 061102 (2016) [ arXiv:1602.03837].
\bibitem{Vishveshwara} C. Vishveshwara  Scattering of gravitational radiation by a schwarzschild Black-hole. \textsl{Nature} {\bf 227}, 936 (1970), DOI 10.1038/227936a0.
\bibitem{Nollert} H. P. Nollert, Quasinormal modes: the characteristicsound'of black holes and neutron stars, \textsl{Class. Quant. Grav.} {\bf 16}, R159 (1999).
\bibitem{Kokkotas}  K. D. Kokkotas, and B. G. Schmidt, Quasi-normal modes of stars and black holes, \textsl{Living Rev. Rel.} {\bf 2}, 2 (1999), [arXiv:gr-qc/9909058].
\bibitem{Berti} E. Berti, V. Cardoso, and  A. O. Starinets, Quasinormal modes of black holes and black branes,\textsl{ Class. Quant. Grav.} {\bf 26}, 163001 (2009), [arXiv:0905.2975].
\bibitem{Konoplya} R. A. Konoplya and A. Zhidenko, Quasinormal modes of black holes: from astrophysics to string theory, \textsl{Rev. Mod. Phys.}, {\bf 83}, 793 (2011).
\bibitem{Danzmann} K. Danzmann, LISA: An ESA cornerstone mission for a gravitational wave observatory, \textsl{Class. Quant. Grav.} {\bf 14}, 1399 (1997).
\bibitem{LISA} LISA collaboration, Laser interferometer space antenna, arXiv:1702.00786.
\bibitem{Hu} W.-R. Hu and Y.-L. Wu, The Taiji program in space for gravitational wave physics and the nature of gravity, \textsl{Natl. Sci. Rev.} {\bf 4}, 685 (2017) .
\bibitem{TianQin} TianQin collaboration, TianQin: a space-borne gravitational wave detector, \textsl{Class. Quant. Grav.} {\bf 33},  035010 (2016), [arXiv:1512.02076].
\bibitem{Bourg} P. Bourg, R. P. Macedo, A. Spiers, B. Leather, B. Bonga, and A. Pound, Quadratic quasinormal mode dependence on linear mode parity, \textsl{Phys. Rev. Lett.} {\bf 134}, 061401 (2025).
\bibitem{Carroll} S. M. Carroll, Spacetime and geometry: an introduction to general relativity, \textsl{Living Rev. Rel.} {\bf 4}, 1 (2001).
\bibitem{Bertone} G. Bertone, D. Hooper, and J. Silk, Particle dark matter: evidence, candidates and constraints, \textsl{Phys. Rep.} {\bf 405}, 279 (2005).
\bibitem{Grumiller} D. Grumiller, Model for gravity at large distances, \textsl{Phys. Rev. Lett.} {\bf 105}, 211303 (2010).
\bibitem{Li_a} Z.-H. Li, X.-J. Wang, L.-Q. Mi, and J.-J. Du, Analysis of the wave equations for the near horizon static isotropic metric, \textsl{Phys. Rev. D} {\bf 95}, 085017 (2017).
\bibitem{Teukolsky} S. A. Teukolsky, Perturbations of a rotating black hole. I. Fundamental equations for gravitational, electromagnetic, and neutrino-field perturbations, \textsl{Astrophys. J.} {\bf 185} 635 (1973).
\bibitem{Torres} G. F. Torres del Castillo, Rarita-Schwinger fields in algebraically special vacuum space-times, \textsl{J. Math. Phys.} {\bf 30} 446 (1989).
\bibitem{Li_b} Z.-H. Li, Unified equation for massless spin particles and new spin coefficient definitions, arXiv:2504.02592.
\bibitem{Ronveaux} A. Ronveaux, \textsl{Heun's Differential Equations}, 1st edn. (Clarendon Press, Oxford, 1995).


\end{thebibliography}
\end{document}